\documentclass[11pt]{JHEP3}

\usepackage{amsmath,latexsym,amssymb}
\usepackage{psfrag}
\usepackage{subfigure}
\usepackage{nicefrac}
\usepackage{bbm}
\usepackage[stable]{footmisc}
\usepackage{fancyhdr}
\usepackage[pdftex]{graphicx}
\usepackage{psfrag}
\usepackage{fancybox}
\usepackage{ifthen}
\usepackage{epsfig}
\usepackage[errorshow]{tracefnt}
\usepackage{multirow}
\usepackage{slashed}

\newcommand\bbone{\ensuremath{\mathbbm{1}}}

\newcommand\beal{\begin{align}}

\newcommand\nn{\nonumber}

\newcommand{\eq}[1]{\begin{equation}#1\end{equation}}

\newcommand{\spl}[1]{\begin{split}#1\end{split}}

\newcommand\benu{\begin{enumerate}}
\newcommand\eenu{\end{enumerate}}
\newcommand\bit{\begin{itemize}}
\newcommand\eit{\end{itemize}}

\newcommand\cm{\mathcal{M}}

\renewcommand{\d}{\ensuremath{\textnormal{d}}}
\renewcommand{\i}{\ensuremath{\textnormal{i}}}

\title{Generalized geometry, calibrations and supersymmetry in diverse dimensions}

\author{Dieter L\"{u}st${}^{\diamondsuit\clubsuit}$, Peter Patalong${}^{\clubsuit}$
 and Dimitrios Tsimpis${}^{\diamondsuit}$ \\

  \begin{itemize}

\item  Arnold-Sommerfeld-Center for Theoretical Physics\\
Department f\"{u}r Physik, Ludwig-Maximilians-Universit\"{a}t M\"{u}nchen\\
Theresienstra\ss e 37, 80333 M\"{u}nchen, Germany
  
\item  Max-Planck-Institut f\"{u}r Physik -- Theorie\\
F\"{o}hringer Ring 6, 80805 M\"{u}nchen, Germany
  \end{itemize}

\bigskip
 E-mail:
\email{luest@mppmu.mpg.de,luest@lmu.de}~,~ \email{peter.patalong@physik.uni-muenchen.de} \& \email{dimitrios.tsimpis@lmu.de }}

\abstract{We consider type II backgrounds of the form $\mathbb{R}^{1,d-1}\times\cm_{10-d}$ for even $d$, preserving 
$2^{d/2}$ real supercharges; for $d=4,6,8$ this is minimal supersymmetry in $d$ dimensions, while for $d=2$ it is $\mathcal{N}=(2,0)$ supersymmetry in two dimensions. For $d=6$ we prove, by explicitly solving the Killing-spinor equations, that there is a one-to-one correspondence between 
background supersymmetry equations in pure-spinor form and D-brane 
generalized calibrations; this correspondence had been known to hold in the $d=4$ case. Assuming the correspondence to hold for all $d$, we list the calibration forms for all admissible D-branes, as well as the background supersymmetry equations in pure-spinor form. We find a number of general features, including the following: The pattern of codimensions at which each calibration form appears exhibits a (mod 4) periodicity. In all cases one of the pure-spinor equations implies that the internal manifold is generalized Calabi-Yau. Our results are manifestly invariant under generalized mirror symmetry.
}

\preprint{LMU-ASC 80/10\\MPP-2010-138}
\begin{document}

\section{Introduction}

Generalized geometry \cite{hitch,gual} (see \cite{koer} for a review) provides a natural mathematical framework for the description of type II flux backgrounds. It has lead to important insights into many recent developments, such as explicit supersymmetric solutions, effective actions, sigma models, as well as supersymmetry breaking and non-geometry. In the language of generalized geometry the supersymmetry conditions for a background of the form $\mathbb{R}^{1,3}\times\cm_6$ are expressed as a set of first-order differential equations for two complex pure spinors of Cliff$(6,6)$ \cite{gran}; the latter can be thought of equivalently as polyforms on $\cm_6$.

The close connection between background supersymmetry and calibrated branes 
\cite{cala,calb,calc} has been noted in various different setups \cite{ma, gauntlett, mb}, and 
calibrations have a natural interpretation within the context of 
generalized geometry \cite{calkoer, calmart, kt,koermart, mk}. For type II backgrounds of the form $\mathbb{R}^{1,3}\times\cm_6$ in particular, this connection works as follows \cite{calmart}: When written in their pure-spinor form, the supersymmetry equations of the background\footnote{In this paper we assume that the internal manifold admits a pair of compatible, globally-defined, nowhere-vanishing pure spinors; as will be reviewed in the following, this implies the reduction of the structure group of the direct sum of the tangent and cotangent bundles of the internal manifold to $SU(k)\times SU(k)$, where $k:=(10-d)/2$. We will moreover assume that the background admits calibrated D-branes; this implies a certain restriction on the norm of the Killing spinors of the background.} are in one-to-one correspondence with the differential conditions obeyed by the calibration forms of all admissible static, magnetic D-branes in that background. 
It is natural to expect that this correspondence extends more generally to 
all type II backgrounds of the form $\mathbb{R}^{1,d-1}\times\cm_{10-d}$ which can be described with generalized geometry (as we will see in the following this requirement restricts $d$ to be even).

In the present paper we show that this is indeed the case for minimally-supersymmetric (four complex supercharges), type II backgrounds of the form $\mathbb{R}^{1,5}\times\cm_{4}$. We prove this by a brute-force computation involving the following steps: a) we explicitly give the general solution of the Killing-spinor equations ({\it i.e.} the supersymmetry conditions) of the background; b) we write down the set of differential equations, in pure-spinor form, obeyed by all admissible static, magnetic, calibrated D-branes in that background; c) we show that the solution of the set of equations in b) is the same as the solution in a).

Based on these results, we conjecture that the one-to-one correspondence between calibrated D-branes and background supersymmetry holds for all (even) $d$, for backgrounds of the form $\mathbb{R}^{1,d-1}\times\cm_{10-d}$ with $2^{d/2-1}$ complex supercharges. This is the amount of supersymmetry parameterized by a complexified Weyl spinor in $d$ dimensions: for $d=4,6,8$ it corresponds to minimal supersymmetry in $d$ dimensions; for $d=2$ it corresponds 
to $\mathcal{N}=(2,0)$ supersymmetry in two dimensions. 

\begin{figure}[h!]\label{table}
\includegraphics{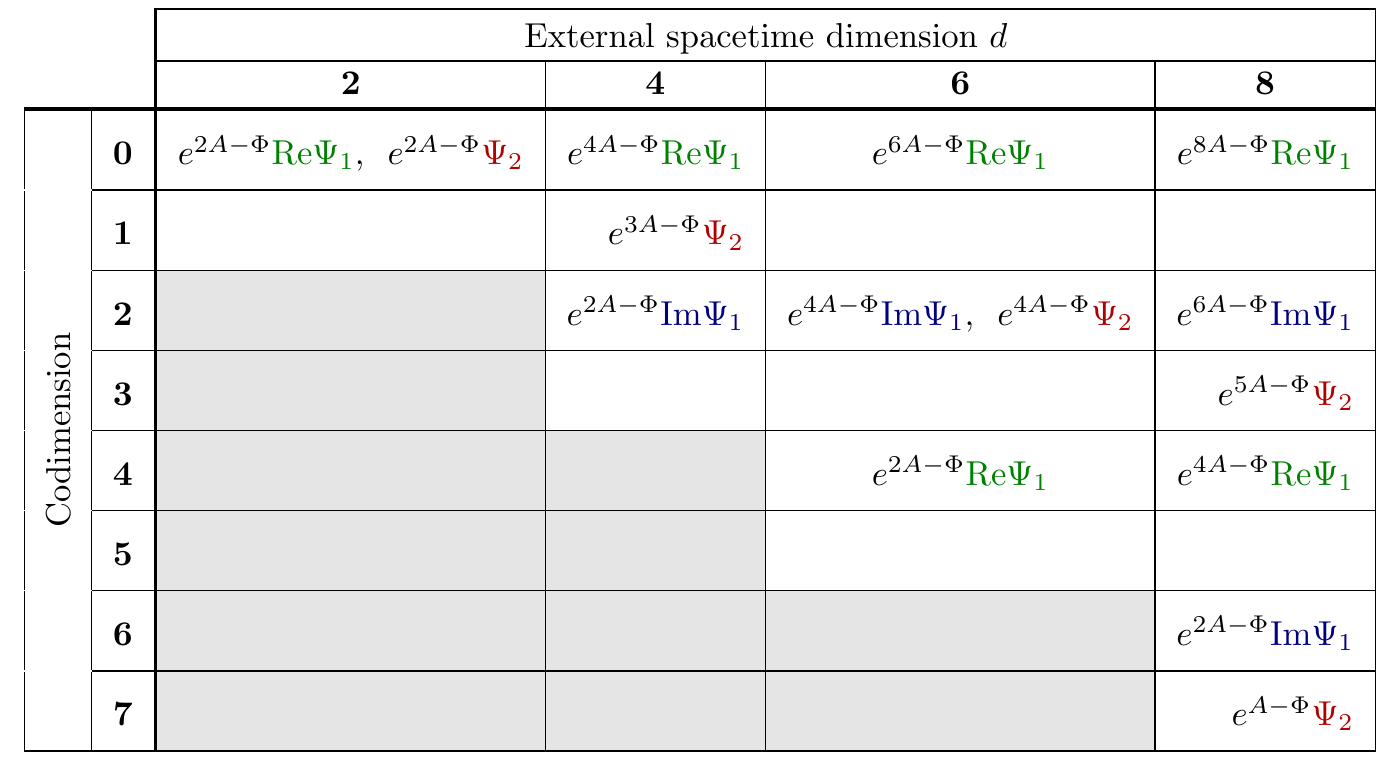}
\caption{The `periodic table' of calibration forms for all $d$, each of them corresponding to a background supersymmetry pure-spinor equation. This one-to-one correspondence had been known to hold in the $d=4$ case, and in the present paper is also shown to hold in the $d=6$ case. Based on these results we conjecture it to hold for the $d=2,8$ cases as well.}
\end{figure}

Assuming the correspondence to be true allows us to deduce the supersymmetry equations for the background in pure-spinor form for the remaining two non-trivial cases corresponding to $d=2,8$,  by performing the much easier task of computing the calibration forms of  all admissible D-branes in that background. 
The summary of the supersymmetry equations for all $\mathbb{R}^{1,d-1}\times\cm_{10-d}$ backgrounds with $2^{d/2-1}$ complex supercharges is given in eq.~(1.1) below; 
\newpage
\begin{figure}[h!]
\begin{flushright}
\includegraphics{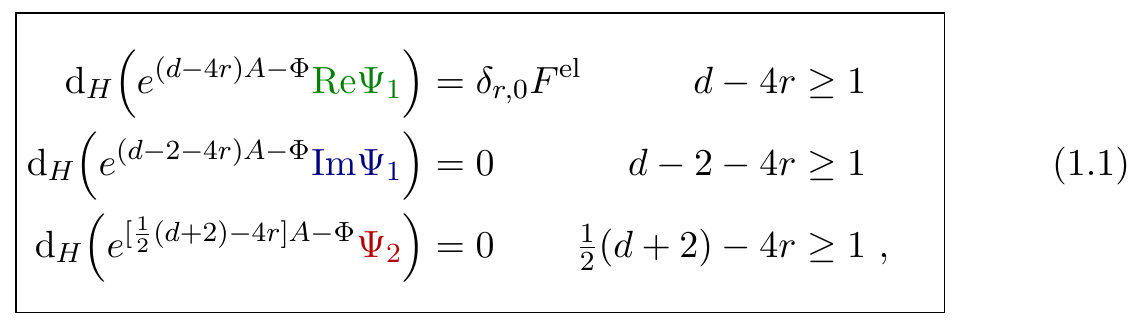}
\end{flushright}
\end{figure}
where $r\in\mathbb{N}$; our notation and conventions for the fluxes are described in detail in section \ref{sec:susyback}. As will be explained in detail in the following sections, each of these pure-spinor equations can be thought of as the differential condition obeyed by the calibration form for a D-brane of the corresponding codimension -- the latter being equal to $d$ minus the coefficient of $A$ in the exponential. 
For each $d$ the different calibration forms  and their corresponding codimensions are given in table \ref{table}.

\renewcommand*\figurename{Table}

Having the complete `periodic table' of pure-spinor supersymmetry equations and their one-to-one correspondence with calibrations, allows one to identify a number of general patterns:
\begin{itemize}

\item The `critical dimension' where there are four (real) pure-spinor equations for two (complex) pure spinors $\Psi_{1,2}$ is $d=4$. For $d>4$ there are 
more equations; for $d<4$ there are fewer equations.

\item One of the equations is always the (twisted) closure of a pure spinor ($\Psi_2$ with an appropriate warp factor). As  will be reviewed in the following, this implies that the internal manifold is Generalized Calabi-Yau.\footnote{In the limit of vanishing flux the internal manifold reduces to an ordinary Calabi-Yau. Recall that in two real dimensions a Calabi-Yau manifold is a $T^2$, while in four real dimensions it is a K3 surface.}

\item $\mathrm{Re}\Psi_1$ is associated with calibrations of codimension 0 mod 4, while $\mathrm{Im}\Psi_1$ is associated with calibrations of codimension 2 mod 4; $\Psi_2$ is associated with calibrations of codimension 
$(d/2-1)$  mod 4.

\item Eqs.~(1.1) as well as the generalized calibrations of table \ref{table} take the same form in both IIA and IIB, hence our results are manifestly invariant under `generalized mirror symmetry'.\footnote{Generalized mirror symmetry may be thought of as the action of reversing the chirality of the pure spinors $\Psi_{1,2}$  in IIA/IIB. Explicitly, if we define $\Psi_+=\Psi_{2/1}$ and $\Psi_-=\Psi_{1/2}$ 
in IIA/IIB, then generalized mirror symmetry acts by exchanging $\Psi_+\leftrightarrow\Psi_-$ and IIA $\leftrightarrow$ IIB. We refer to \cite{koer} for further discussion.}

\end{itemize}
Finally, as an illustration of the pure-spinor formalism in the $d=6$ case, we construct a type IIB warped K3 solution with spacetime-filling D5 branes  localized on the K3. In the degenerate limit where K3 is replaced by a $T^4$ we show that the solution coincides with the one obtained using the `harmonic superposition rules' for a stack of D5 branes in flat space. We also construct a T-dual IIA warped $S^1\times T^3$ solution with spacetime-filling D6 branes  localized on the $T^3$ and wrapping the $S^1$.

The remainder of the paper is organized as follows. In  section \ref{sec:generalized} we give a brief introduction to generalized geometry. The type II flux backgrounds which we consider are described in detail in section \ref{sec:susyback}. Section \ref{sec:calib} contains a review of calibrations in the present context. The different admissible D-brane calibrations for all $d$ are constructed in section \ref{sec:diverse}. We give our conclusions in section \ref{sec:conclusions}.  In appendix \ref{sec:spinors} we list some useful spinor and gamma-matrix identities. Appendix \ref{sec:explicit} contains our proof of the one-to-one correspondence between background supersymmetry equations and calibrations in $d=6$. The warped K3 and $S^1\times T^3$ solutions are 
given in appendix \ref{sec:warpedk3}.

\section{Generalized geometry}\label{sec:generalized}

For completeness we briefly review here the relevant concepts of generalized complex geometry \cite{hitch,gual}. We refer to  {\it e.g.}  the recent review \cite{koer} for detailed explanations and references.

 {\it Generalized almost complex structures}

Generalized complex geometry is an extension of both complex and symplectic geometry, interpolating, in a sense which we will make precise in the following,  between these two special cases. Consider an even-dimensional 
manifold $\cm_{2k}$. One can equip the sum of tangent and cotangent bundles $T\oplus T^*$ with a metric of maximally indefinite signature $\mathcal{G}$ (the pairing between vectors and forms), reducing the structure group to $O(2k,2k)$. Imposing in addition the existence of an almost complex structure $\mathcal{I}$ on $T\oplus T^*$ associated with the metric $\mathcal{G}$ ({\it i.e.} such that $\mathcal{G}$ is hermitian with respect to $\mathcal{I}$), further reduces the structure group to $U(k,k)$. 

 A pair $\mathcal{I}_{1,2}$ of {\it compatible} almost complex structures on $T\oplus T^*$ ({\it i.e}. such that they commute and they give rise to a positive definite metric) further reduces the structure group to $U(k)\times U(k)$. The metric on $T\oplus T^*$ associated with the pair $\mathcal{I}_{1,2}$ can be seen to give rise to both a positive definite metric $g$ and a $B$-field on $T$.

 {\it (Generalized) almost complex structures and pure spinors}

Just as there is an equivalence between almost complex structures on $T$ and line bundles of pure Weyl spinors of $\mathrm{Cliff}(2k)$,\footnote{Recall that 
pure Weyl spinors may be defined as the spinors which are annihilated by precisely those gamma matrices that are holomorphic (or antiholomorphic, depending on the convention) with respect to an almost complex structure.} there is an equivalence between almost complex structures on $T\oplus T^*$ and line bundles of pure spinors of $\mathrm{Cliff}(2k,2k)$. Demanding that the line bundle 
of pure spinors of $\mathrm{Cliff}(2k,2k)$ have a global section, reduces the structure group of $T\oplus T^*$ from $U(k,k)$ (which was accomplished by the existence of a generalized almost complex structure) to $SU(k,k)$.

 {\it Spinors on $T\oplus T^{*}$,  bispinors on $T$, polyforms  in $\Lambda^\bullet T^*$}

There is a natural action of $T\oplus T^{*}$ on the bundle $\Lambda^\bullet T^*$ 
of differential forms on $\cm_{2k}$, whereby every vector acts by contraction and every one-form by exterior multiplication. It can easily be seen that this action obeys the Clifford algebra $\mathrm{Cliff}(2k,2k)$ associated with the maximally indefinite metric $\mathcal{G}$ on $T\oplus T^*$. It follows that there is an isomorphism $\mathrm{Cliff}(2k,2k)\thickapprox {\rm End}(\Lambda^\bullet T^*)$, which means that spinors on $T\oplus T^{*}$ can be identified with {\it polyforms} ({\it i.e.} sums of forms of different degrees) in $\Lambda^\bullet T^*$.

On the other hand, there is a correspondence between polyforms of $\Lambda^\bullet T^*$ 
and {\it bispinors} on $T$. This correspondence is a canonical isomorphism, 
up to a choice of the volume form, and is explicitly realized by the {\it Clifford map}:
\eq{\label{cliffmap}
\psi_\alpha\otimes\widetilde{\chi}_\beta=\frac{1}{2^k}\sum_{p=0}^{2k}\frac{1}{p!}
(\widetilde{\chi}\gamma_{m_p\dots m_1}\psi)~\!\gamma^{m_1\dots m_p}_{\alpha\beta}
\longleftrightarrow
\frac{1}{2^k}\sum_{p=0}^{2k}\frac{1}{p!}
(\widetilde{\chi}\gamma_{m_p\dots m_1}\psi)~\!e^{m_1}\wedge\dots\wedge e^{m_p}
~,}
where the first equality is the Fierz identity. 

 {\it Pairs of compatible pure spinors and $SU(k)\times SU(k)$ structures}

It follows from the above discussion that the condition of compatibility  
of a pair of generalized almost complex structures should be expressible as 
a condition of compatibility on a pair of (line bundles of) pure spinors of $\mathrm{Cliff}(2k,2k)$ -- which, as already mentioned,  can 
alternatively be thought of as either bispinors of $\mathrm{Cliff}(2k)$ or,  
through eq.~(\ref{cliffmap}), as polyforms. Indeed, the most general 
form of a pair $\Psi_{1,2}$ of compatible pure spinors  of $\mathrm{Cliff}(2k,2k)$ is given by:
\eq{\spl{\label{purespinors}
\Psi_1&=\frac{(2i)^{k}}{|a|^2}\eta_1\otimes\widetilde{\eta^c_2}\\
\Psi_2&=\frac{(2i)^{k}}{|a|^2}\eta_1\otimes\widetilde{\eta}_2
~,}}
where $\eta_{1,2}$ are pure spinors\footnote{\label{f1} Note that 
for $k\leq3$, Weyl spinors of $\mathrm{Cliff}(2k)$ are automatically pure. 
For the case $k=4$ one has to impose in addition 
one complex condition; we will return to this in section \ref{d2}.} of $\mathrm{Cliff}(2k)$. 
The normalization above is chosen for future convenience, and we have 
imposed that the background admits calibrated branes, in which case $\eta_{1,2}$ have equal norm: $|a|^2:=\widetilde{\eta}_1\eta_1^c=\widetilde{\eta}_2\eta_2^c$; see appendix \ref{sec:spinors} for our spinor conventions.

Provided the pair of pure spinors above is globally defined and nowhere vanishing (in other words: if the corresponding line bundles of pure spinors have nowhere-vanishing global sections), the structure group of $T\oplus T^*$ is further reduced from $U(k)\times U(k)$ (which was accomplished by the existence of a pair of compatible generalized almost complex structures) to $SU(k)\times SU(k)$.

 {\it Generalized complex manifolds, and GCY}

The correspondence between generalized almost complex structures 
and pure spinors allows one to express the condition of integrability of a 
generalized almost complex structure as a certain first-order differential equation 
for the associated pure spinor, which may then also be called integrable. A manifold $\cm_{2k}$ is called {\it generalized complex} if it admits an integrable pure spinor. It can be shown that if $\cm_{2k}$ is generalized complex, it is locally equivalent to $\mathbb{C}^q\times(\mathbb{R}^{2(k-q)},J)$, with $J$ the standard 
symplectic structure; thus generalized complex geometry can be said to be an interpolation between complex and symplectic geometries. The integer $q$ is called the {\it type}, and need not be constant over $\cm_{2k}$.

A {\it generalized Calabi-Yau (GCY)} is a special case of a generalized complex manifold. It is defined as a manifold $\cm_{2k}$ on which a pure spinor $\Psi$ exists, obeying the differential condition\footnote{This is also sometimes called the `twisted' Calabi-Yau condition; the pure spinor 
$\Psi$ is thought of as a polyform in $\Lambda^\bullet T^*$ via the 
Clifford map (\ref{cliffmap}).}
\eq{\label{gcy}
\mathrm{d}_H\Psi=0~,}
where $\mathrm{d}_H:=d+H\wedge~\!$ and $H=dB$ is the field strength of the $B$ field. The presence of the 
latter should not be too surprising, as we have already mentioned that pairs of compatible pure spinors naturally incorporate a $B$ field.

\section{Supersymmetric flux backgrounds}\label{sec:susyback}

Let us now describe our supergravity setup in more detail. 

We consider ten-dimensional type IIA/IIB backgrounds of the form:
\eq{\label{10dmetric}
\d s^2 = e^{2A}\d s^2(\mathbb{R}^{1,d-1})+\d s^2(\cm_{10-d})  
~,}
where: 
\eq{
d=2,4,6,8
~.} 
The case $d=10$ is trivial and will 
not be considered separately. The warp factor $A$ is taken to only depend on 
the coordinates of the `internal' Riemannian manifold $\cm_{10-d}$.

We assume that not all 
RR charges are zero; the case with zero RR charges has already 
been analyzed in \cite{gauntlett}. The most general 
RR charges respecting the Poincar\'e symmetry of our setup are of the form:\footnote{We follow the `democratic' supergravity conventions of \cite{march}, see appendix A therein. }
\eq{\label{fluxan}
F^{\mathrm{tot}}=\mathrm{vol}_d\wedge F^{\mathrm{el}}+F
~,}
where $\mathrm{vol}_d$ is the unwarped volume element of $\mathbb{R}^{1,d-1}$, 
 and we are using polyform notation. 
We denote by $F$ the `magnetic' RR charges with legs 
on the intenral space $\cm_{10-d}$. 
The ten-dimensional Hodge duality relates $F$ to the `electric' RR charges 
via:
\eq{
F^{\mathrm{el}}=\left(e^{A}\right)^d\star_{10-d}\sigma(F)
~,}
where the Hodge star above is with respect to the internal metric, and 
the involution $\sigma$ acts by inverting the order of the form indices. 

We consider backgrounds preserving $2^{d/2-1}$ complex supercharges. 
Note that the dimension of a Weyl spinor of $\mathbb{R}^{1,d-1}$ is precisely $\mathrm{dim}(\mathrm{Weyl}_d)=2^{d/2-1}$, so that 
the supercharges are parameterized by a complexified\footnote{We use the terminology `complexified' 
for a Weyl spinor with complex components. The term `complex Weyl spinor' 
is reserved for Weyl spinors whose complex conjugate has opposite chirality.}  Weyl spinor $\zeta$ 
of $\mathbb{R}^{1,d-1}$. More explicitly, 
the Killing spinors of the ten-dimensional background are given by:
\eq{\label{spindecompa}
\epsilon_i=\zeta\otimes\eta_i+\mathrm{c.c.} 
~,}
where $i=1,2$, so that $\epsilon_{1,2}$ are ten-dimensional Majorana-Weyl spinors of opposite, the same chirality for IIA, IIB respectively. The spinors 
$\eta_{1,2}$ are pure Weyl spinors ({\it cf}. footnote \ref{f1}) of $\mathrm{Cliff}(10-d)$ of opposite, the same chirality for IIA, IIB respectively. The precise form of the complex 
conjugate on the right hand side of the equation above depends on the dimension $d$ and will be given explicitly in the following.

For $d=4,6,8$ the Killing spinor ansatz given in eq.~(\ref{spindecompa}) corresponds to minimal supersymmetry in $d$ dimensions; for $d=2$ it corresponds 
to $\mathcal{N}=(2,0)$ supersymmetry in two dimensions.

\subsection{Calibrations}\label{sec:calib}

The close connection between supersymmetry and  calibrations was noted some time ago \cite{cala,calb,calc}. More recently, generalized calibrations in flux backgrounds were shown to have a natural interpretation 
in terms of generalized geometry \cite{calkoer,calmart,koermart}. In this section we will briefly review the relevant results, referring the reader to \cite{koer} or the original literature for further details. 

Consider the energy density $\mathcal{E}(\Sigma,\mathcal{F})$ of a static, magnetic ({\it i.e.} without electric worldvolume flux) D-brane in our setup, filling $q$ external spacetime dimensions and  wrapping a cycle $\Sigma$ in the internal space (for our purposes it will not be necessary to take higher-order corrections into consideration):
\eq{\label{endens}
\mathcal{E}(\Sigma,\mathcal{F})
=
e^{qA-\Phi}\sqrt{det(g+\mathcal{F})}-\delta_{q,d}
\left(C^{\mathrm{el}}\wedge e^\mathcal{F}\right)_\Sigma
~,}
where $g$ is the induced worldvolume metric on $\Sigma$, $\mathcal{F}$ is the worldvolume flux: $d\mathcal{F}=H|_\Sigma$, and  $C^{\mathrm{el}}$ is the electric RR flux potential: $\mathrm{d}_HC^{\mathrm{el}}=F^{\mathrm{el}}$, {\it cf}. eq.~(\ref{fluxan}). Note that unless the brane fills all the external spacetime directions, the second term on the right hand side above vanishes. This property of the energy density follows from the form of the ansatz for the RR fields,  eq.~(\ref{fluxan}), which is such that it preserves the $d$-dimensional Poincar\'e invariance of the background. 

A polyform $\omega$ (defined in the whole of the internal space) is a {\it generalized calibration form} if, for any cycle $\Sigma$, it satisfies the algebraic inequality:
\eq{\label{calalg}
\left(\omega\wedge e^{\mathcal{F}}\right)_\Sigma\leq \mathrm{d}\sigma e^{qA-\Phi}\sqrt{det(g+\mathcal{F})}
~,}
where $\sigma$ collectively denotes the coordinates of $\Sigma$, together with the differential condition:\footnote{
Alternatively the calibration form is sometimes defined to obey:
\eq{
\left(\omega'\wedge e^{\mathcal{F}}\right)_\Sigma\leq \mathrm{d}\sigma\mathcal{E}(\Sigma,\mathcal{F})
~,\nn}
as well as the differential condition:
\eq{
\mathrm{d}_H\omega'=0
~.\nn
}
The two definitions are related by: $\omega'=\omega-\delta_{q,d}C^{\mathrm{el}}$; the one we adopt in the main text is more natural from the point of view of the calibrations/background supersymmetry correspondence.}
\eq{\label{caldiff}
\mathrm{d}_H\omega=\delta_{q,d}F^{\mathrm{el}}
~.}
A generalized submanifold $(\Sigma,\mathcal{F})$ is called {\it calibrated by $\omega$}, if it saturates the bound given in eq.~(\ref{calalg}) above.

The upshot of the above discussion is that D-branes wrapping generalized calibrated submanifolds minimize their energy within their (generalized) homology class. Recall that $(\Sigma,\mathcal{F})$,  $(\Sigma',\mathcal{F}')$ are in the same generalized homology class if there is a cycle $\widetilde{\Sigma}$ such that $\partial\widetilde{\Sigma}=\Sigma'-\Sigma$ 
and there exists an extension of the worldvolume flux $\widetilde{\mathcal{F}}$ on $\widetilde{\Sigma}$ such that: $\widetilde{\mathcal{F}}|_\Sigma=\mathcal{F}$ and  $\widetilde{\mathcal{F}}|_{\Sigma'}=\mathcal{F}'$. 
Then, if $(\Sigma,\mathcal{F})$ is calibrated by $\omega$ we have, using 
Stokes theorem as well as eqs.~(\ref{endens}-\ref{caldiff}):
\eq{
\int_{\Sigma'} \mathrm{d}\sigma~\! \mathcal{E}(\Sigma',\mathcal{F}')\geq
\int\left(\omega-\delta_{q,d}C^{\mathrm{el}}\right)_{\Sigma'}\wedge e^{\mathcal{F}'}
=\int \left(\omega-\delta_{q,d}C^{\mathrm{el}}\right)_\Sigma\wedge e^\mathcal{F}
=\int_\Sigma \mathrm{d}\sigma ~\! \mathcal{E}(\Sigma,\mathcal{F})
~.}

For type II backgrounds, the generalized calibration form $\omega$ can be constructed explicitly as follows. As explained in \cite{kt} one has to break the 
$SO(1,9)$ symmetry of the tangent bundle of spacetime to  $SO(9)$. We
 decompose the Killing spinors:
\eq{\label{spindecompb}
\epsilon_1 = \left(\begin{array}{c} 1 \\ 0 \end{array}\right) \otimes
\chi_1  \, , \qquad
\epsilon_2 = \left(\begin{array}{c} 1 \\ 0 \end{array}\right) \otimes
\chi_2  \quad \text{(IIB)} \, , \quad
\epsilon_2 = \left(\begin{array}{c} 0 \\ 1 \end{array}\right) \otimes
\chi_2 \quad \text{(IIA)} \, ,
}
where $\chi_{1,2}$ are real, commuting spinors of $SO(9)$. The ten-dimensional gamma matrices decompose accordingly as
\eq{
\label{epsdecomp}
\qquad \Gamma_{{0}}=
(i \sigma_2) \otimes \bbone \, ,\qquad\Gamma_{{m}} = \sigma_1 \otimes {\gamma}_{{m}} \, ,  \qquad \Gamma_{11} =  \sigma_3
\otimes \bbone~,
}
where $\sigma_i$ are the Pauli matrices, 
and ${\gamma}_{{m}}=e_m{}^a\gamma_a$ are 
nine-dimensional gamma matrices, with $e_m{}^a$ the 
(warped) vielbein associated with the metric in (\ref{10dmetric}).

Using the $SO(9)$ spinors $\chi_{1,2}$ one can construct on the nine-dimensional space the real polyform 
\eq{
\Omega:= \sum_{p\ \text{even/odd}} \frac{e^{A-\Phi}}{p!|a|^2}
\left(\widetilde{\chi}_1 {\gamma}_{m_1 \ldots m_p} \chi_2\right)
\; \mathrm{d} x^{m_1}\wedge \ldots\wedge \mathrm{d} x^{m_p} \, ,
\label{calform}
}
where one has to sum over $p$ even/odd in IIA/IIB respectively, and we have normalized:
\eq{\label{chinorm}
\widetilde{\chi}_1\chi_1=\widetilde{\chi}_2\chi_2=|a|^2~.
}
The equality of the norms of $\chi_{1,2}$ can be seen to follow from the requirement that the background admits kappa-symmetric branes which do not break the supersymmetry of the background. In this case, it 
can be shown on rather general grounds that supersymmetry implies $|a|^2\propto e^A$. We will choose the proportionality constant so that:
\eq{\label{ea}
|a|^2= e^A
~.}

Let us denote by $\Omega^{(q-1)}$ the sum of all terms in (\ref{calform}) which contain exactly $(q-1)$ external spatial directions. Slightly adapting the proof in appendix A.3 of \cite{koermart} to the present setup, 
it can then be seen that the following polyform: 
\eq{\label{calfin}
\omega^{(d-q)}:=\frac{\Omega^{(q-1)}}{\mathrm{vol}_{\mathrm{sp}}^{(q-1)}}~, 
}
is a calibration form for static, magnetic D-branes filling $q$ external spacetime directions.  In the above $\mathrm{vol}_{\mathrm{sp}}^{(q-1)}$ is the unwarped volume density along the $(q-1)$ external spatial directions that the brane fills, and the superscript of $\omega$ denotes the codimension with respect to the external $d$-dimensional  spacetime. Note that this is {\it not} in general  equal to the codimension of the branes with respect to the ten-dimensional spacetime: the branes wrap $p$-dimensional 
cycles in the internal space such that $p+q=$ odd/even in IIA/IIB.

Indeed it can be seen that $\omega^{(d-q)}$ defined in eq.~(\ref{calfin}) satisfies the algebraic inequality (\ref{calalg}). 
Moreover, for $\omega^{(d-q)}$ to satisfy the differential condition (\ref{caldiff}), it suffices that \cite{koermart}:
\eq{\label{calsuff}
\iota_{v_+}F=0~;~~\mathrm{and}~~v_-=0
~,}
where the vectors $v_\pm$ are given by:
\eq{\label{vdef}
v^m_\pm:=
\left\{
\begin{array}{ll}
(\widetilde{\chi}_1\gamma^m\chi_1) \mp (\widetilde{\chi}_2\gamma^m\chi_2) ~,& ~~~\mathrm{in~IIA}\\
(\widetilde{\chi}_1\gamma^m\chi_1) \pm (\widetilde{\chi}_2\gamma^m\chi_2)~,& ~~~\mathrm{in~IIB}\\
\end{array}
\right.
~.}
It can be easily verified that the conditions in eq.~(\ref{calsuff}) are automatically satisfied for all the backgrounds which will be considered in sections \ref{d4} - \ref{d2}.

\section{Calibrations and supersymmetry in diverse dimensions}\label{sec:diverse}

We will now apply the method described in section \ref{sec:calib} to 
construct the calibration forms for all supersymmetric
 backgrounds of the type described at the beginning of section \ref{sec:susyback}.

We start by reviewing the well-known $d=4$ case 
 in section \ref{d4}. The supersymmetry equations can be cast in the form 
of four real first-order differential equations for two (complex) pure spinors 
of Cliff$(6,6)$, one of which imposes the GCY condition \cite{gran}.
Moreover, the result of the calibration analysis is that there is a one-to-one correspondence between the supersymmetry pure-spinor equations 
for the background (assuming the background admits calibrations so that   eq.~(\ref{chinorm}) holds), and the differential equations obeyed by the generalized calibrations in that background \cite{calmart}. 

We then repeat the analysis for the $d=6$ case in section \ref{d6}. We construct the generalized calibrations for the background and we express the differential equations which they obey as a set of five real first-order equations for two (complex) pure spinors of Cliff$(4,4)$. A brute force calculation given in appendix \ref{sec:explicit} shows that, as for the $d=4$ case, the content of these five real pure-spinor equations is precisely equivalent to the supersymmetry equations for the background -- assuming  the background admits calibrations. Moreover, as in the $d=4$ case, one of the consequences of supersymmetry is that the internal manifold is GCY.

The remaining two cases, $d=8,2$, are discussed in sections \ref{d8}, \ref{d2} respectively. We work out the differential conditions for the generalized calibrations in these backgrounds and express them as first-order differential equations for two (complex) pure spinors. As in the previous two cases, the equations imply that the internal manifold is GCY. For these last two cases we do not verify that the differential conditions thus obtained are equivalent to the superymmetry equations -- although we conjecture it to be true, based on the results of the $d=4,6$ cases. The result of the analysis for all $d$ is summarized in eq.~(1.1) and table \ref{table} of the introduction.

Before we proceed to the case-by-case analysis, let us also mention that the 
spinor ansatz of eq.~(\ref{spindecompa}) must be modified in order to take into account the $SO(1,9)\rightarrow SO(9)$ reduction of eq.~(\ref{spindecompb}). Under $SO(1,d-1)\rightarrow SO(d-1)$ the Weyl spinor $\zeta$, which transforms in the ${ 2^{d/2-1}_+}$ of $SO(1,d-1)$, restricts to the ${ 2^{d/2-1}}$ of $SO(d-1)$; we will denote the restriction to  $SO(d-1)$ by $\theta$. The spinor ansatz then takes the form of eq.~(\ref{spindecompb}), with:
\eq{\label{spindecompc}
\chi_i=\frac{1}{\sqrt{2}}\left(\theta\otimes\eta_i+\mathrm{c.c.}\right)
~,}
where $i=1,2$ and $\eta_i$ are the same pure Weyl spinors of Cliff$(10-d)$ as in eq.~(\ref{spindecompa}).  We assume the   normalization: $\widetilde{\theta}\theta^c=1$, $\widetilde{\eta_i}\eta_i^c=|a|^2$, $i=1,2$, so that (\ref{chinorm}) is obeyed. 
The precise form of the complex conjugate on the right-hand sides of the equations above will be given explicitly for each case in the following.

This restriction on the norms of the $\eta_i$'s (following from the requirement that the background admit calibrated D-branes) implies, taking eq.~(\ref{ea}) into account,  that the pair of compatible pure spinors defined in  eq.~(\ref{purespinors}) are non-vanishing provided the warp factor $A$ is finite. We will assume that they are also globally defined; as reviewed in section \ref{sec:generalized}, this implies the reduction of the structure group\footnote{The reduction of the structure group of the internal manifold is, in general, {\it not} a necessary condition for supersymmetric backgrounds \cite{tsimpism}; see \cite{mcorist} for a recent explicit example.} of the direct sum of the tangent and cotangent bundles of the internal manifold to $SU(k)\times SU(k)$, where $k:=(10-d)/2$.

\subsection{d=4}\label{d4}

The pure spinor equations for supersymmetric backgrounds of the form  $\mathbb{R}^{1,3}\times\cm_6$ with two complex supercharges, {\it i.e.} minimal supersymmetry in four dimensions, were worked out 
in \cite{gran}. In \cite{calmart} it was subsequently shown that the differential conditions for generalized calibrations for static, magnetic  D-branes in this background are in one-to-one correspondence with the supersymmetry equations in pure-spinor form. We shall review this case here for completeness.

Under $SO(9)\rightarrow SO(3)\times SO(6)$ the nine-dimensional gamma matrices and  charge conjugation matrix decompose as:
\eq{
\label{epsdecompd4}
\qquad \Gamma_{{i}}=
\sigma_i \otimes \gamma_7 \, ,\qquad\Gamma_{{m+3}} = \bbone \otimes {\gamma}_{{m}} \, ,  \qquad C_{9} =  C_3\otimes \gamma_7C_6~,
}
where $\left\{\sigma_i, ~i=1,2,3\right\}$, $\left\{\gamma_m, ~m=1,\dots,6\right\}$ are three-, six-dimensional gamma matrices, respectively, and $\gamma_7$ is the six-dimensional chirality matrix. 
In our spinor conventions, described in appendix \ref{sec:spinors}, it can then  be seen that the explicit form of the spinor decomposition eq.~(\ref{spindecompc}) reads:
\eq{
\chi_1=\frac{1}{\sqrt{2}}\left(\theta\otimes\eta_1-\theta^c\otimes\eta_1^c\right)~;
~~~~\chi_2=\frac{1}{\sqrt{2}}\left(\theta\otimes\eta_2\pm\theta^c\otimes\eta_2^c\right)
~,}
where $\theta$ is in the {\bf 2} of $SO(3)$ and $\eta_{1}$, $\eta_{2}$ are Weyl spinors in the {\bf 4} of $SO(6)$, with $\gamma_7\eta_1=\eta_1$ and $\gamma_7\eta_2=\mp\eta_2$ in IIA/IIB.

Plugging the above expressions for $\chi_{1,2}$ into eq.~(\ref{calform}, \ref{calfin}), taking eqs.~(\ref{bilins}, \ref{bilinstr}, \ref{bilinscomp}) into account,  we 
find that the only non-vanishing calibration forms occur at codimensions zero (spacetime-filling), one (domain walls), and two (D-strings) with respect to the external spacetime. The explicit expressions read:
\eq{\spl{\label{cald4}
\omega^{(0)}&= e^{4A-\Phi}\mathrm{Re}\Psi_1\\
\omega^{(1)}&= e^{3A-\Phi}\mathrm{Im}\left(e^{i\varphi}\Psi_2\right)\\
\omega^{(2)}&= e^{2A-\Phi}\mathrm{Im}\Psi_1
~,}}
where we have taken 
eq.~(\ref{purespinors}) into account. 
The phase $\varphi$ on the right hand side of the second line comes from the normalization:
\eq{\label{phased4}
\frac{1}{2}(\widetilde{\theta}\sigma_{ij}\theta)~\!\mathrm{d}x^{i}\wedge \mathrm{d}x^j=e^{i\varphi}\mathrm{vol}_{\mathrm{sp}}^{(2)}
~,}
where as in eq.~(\ref{calfin}) $\mathrm{vol}_{\mathrm{sp}}^{(q-1)}$ is  the 
unwarped volume density along the $(q-1)$-dimensional external space that the brane fills.

The connection between the generalized calibrations (\ref{cald4}) and 
background supersymmetry is made by taking the differential equation (\ref{caldiff}) into account. We thus obtain the following equations:
\eq{\spl{\label{susyd4}
\mathrm{d}_H\left( e^{4A-\Phi}\mathrm{Re}\Psi_1\right)&=F^{\mathrm{el}}\\
\mathrm{d}_H\left( e^{2A-\Phi}\mathrm{Im}\Psi_1\right)&=0\\
\mathrm{d}_H\left(e^{3A-\Phi}\Psi_2\right)&=0
~,}}
which are equivalent to the background  supersymmetry equations \cite{gran}. Note that the last equation above is precisely the GCY condition for the pure spinor $e^{3A-\Phi}\Psi_2$; it is obtained by imposing $\mathrm{d}_H\omega^{(1)}=0$ for all $\varphi$, with $\omega^{(1)}$ given in eq.~(\ref{cald4}).

The background supersymmetry equations (\ref{susyd4}) and their correspondence with the D-brane calibrations eq.~(\ref{cald4}) is summarized in the $d=4$ column of table \ref{table} and eq.~(1.1) of the introduction.

\subsection{d=6}\label{d6}

Let us now consider supersymmetric backgrounds of the form  $\mathbb{R}^{1,5}\times\cm_4$ preserving four complex supercharges, {\it i.e.} 
minimal supersymmetry in six dimensions. 
We will show that, as in the $d=4$ case,  the differential conditions for generalized calibrations for static, magnetic D-branes in this background are in one-to-one correspondence with the supersymmetry equations in pure-spinor form.

Under $SO(9)\rightarrow SO(5)\times SO(4)$ the nine-dimensional gamma matrices and  charge conjugation matrix decompose as:
\eq{
\label{epsdecompd6}
\qquad \Gamma_{{i}}=
\sigma_i \otimes \gamma_5 \, ,\qquad\Gamma_{{m+5}} = \bbone \otimes {\gamma}_{{m}} \, ,  \qquad C_{9} =  C_5\otimes C_4~,
}
where $\left\{\sigma_i, ~i=1,\dots,5\right\}$, $\left\{\gamma_m, ~m=1,\dots,4\right\}$ are five-, four-dimensional gamma matrices, respectively, and $\gamma_5$ is the four-dimensional chirality matrix. 
It can then  be seen that the explicit form of the spinor decomposition eq.~(\ref{spindecompc}) reads:
\eq{
\chi_i=\frac{1}{\sqrt{2}}\left(\theta\otimes\eta_i+\theta^c\otimes\eta_i^c\right)
~,}
where $i=1,2$; $\theta$ is in the {\bf 4} of $SO(5)$ and $\eta_{1}$, $\eta_{2}$ are Weyl spinors in the {\bf 2} of $SO(4)$, with $\gamma_5\eta_1=\eta_1$ and $\gamma_7\eta_2=\mp\eta_2$ in IIA/IIB. 

Plugging the above expressions for $\chi_{1,2}$ into eq.~(\ref{calform}, \ref{calfin}), taking eqs.~(\ref{bilins}, \ref{bilinstr}, \ref{bilinscomp}) into account,  we 
find that the only non-vanishing calibration forms occur at codimensions zero (spacetime-filling), two and four with respect to the external spacetime. The explicit expressions read:
\eq{\spl{\label{cald6}
\omega^{(0)}&= e^{6A-\Phi}\mathrm{Re}\Psi_1\\
\omega^{(2)}&=e^{4A-\Phi}\mathrm{Re}\left(e^{i\varphi}\Psi_2\right)+ e^{4A-\Phi}\mathrm{Im}\Psi_1\\
\omega^{(4)}&= e^{2A-\Phi}\mathrm{Re}\Psi_1
~,}}
where  we have taken 
eq.~(\ref{purespinors}) into account. The phase $\varphi$ on the right hand side of the second line comes from the normalization:
\eq{\label{phased6}
\frac{1}{3!}(\widetilde{\theta}\sigma_{ijk}\theta)~\!\mathrm{d}x^{i}\wedge \mathrm{d}x^j\wedge \mathrm{d}x^k=e^{i\varphi}\mathrm{vol}_{\mathrm{sp}}^{(3)}
~,}
where as in eq.~(\ref{calfin}) $\mathrm{vol}_{\mathrm{sp}}^{(q-1)}$ is  the 
unwarped volume density along the $(q-1)$-dimensional external space that the brane fills.

The connection between the generalized calibrations (\ref{cald6}) and 
background supersymmetry is made by taking the differential equation (\ref{caldiff}) into account. We thus obtain the following equations:
\eq{\spl{\label{susyd6}
\mathrm{d}_H\left( e^{6A-\Phi}\mathrm{Re}\Psi_1\right)&=F^{\mathrm{el}}\\
\mathrm{d}_H\left( e^{4A-\Phi}\mathrm{Im}\Psi_1\right)&=0\\
\mathrm{d}_H\left( e^{2A-\Phi}\mathrm{Re}\Psi_1\right)&=0\\
\mathrm{d}_H\left(e^{4A-\Phi}\Psi_2\right)&=0
~.}}
The last equation above is precisely the GCY condition for the pure spinor $e^{4A-\Phi}\Psi_2$; that and the equation in the second line above are obtained by imposing $\mathrm{d}_H\omega^{(2)}=0$ for all $\varphi$, with $\omega^{(2)}$ given in eq.~(\ref{cald6}).

A tedious but straightforward brute-force calculation, given in appendix \ref{sec:explicit}, shows that the content of eqs.~(\ref{susyd6}) is precisely equivalent to the supersymmetry equations for the background, thus proving the one-to-one correspondence between supersymmetry and D-brane calibrations. This correspondence is summarized in the $d=6$ column of table \ref{table} and eq.~(1.1) of the introduction.

\subsection{d=8}\label{d8}

Let us consider supersymmetric backgrounds of the form  $\mathbb{R}^{1,7}\times\cm_2$ preserving eight complex supercharges, {\it i.e.} minimal supersymmetry in eight dimensions.

Under $SO(9)\rightarrow SO(7)\times SO(2)$ the nine-dimensional gamma matrices and  charge conjugation matrix decompose as:
\eq{
\label{epsdecompd8}
\qquad \Gamma_{{i}}=
\sigma_i \otimes \gamma_3 \, ,\qquad\Gamma_{{m+7}} = \bbone \otimes {\gamma}_{{m}} \, ,  \qquad C_{9} =  C_7\otimes \gamma_3C_2~,
}
where $\left\{\sigma_i, ~i=1,\dots,7\right\}$, $\left\{\gamma_m, ~m=1,2\right\}$ are seven-, two-dimensional gamma matrices, respectively, and $\gamma_3$ is the two-dimensional chirality matrix. It can then  be seen that the explicit form of the spinor decomposition eq.~(\ref{spindecompc}) reads:
\eq{
\chi_1=\frac{1}{\sqrt{2}}\left(\theta\otimes\eta_1-\theta^c\otimes\eta_1^c\right)~;~~~
\chi_2=\frac{1}{\sqrt{2}}\left(\theta\otimes\eta_2\pm\theta^c\otimes\eta_2^c\right)
~,}
where $\theta$ is in the {\bf 8} of $SO(7)$ and $\eta_{1}$, $\eta_{2}$ are Weyl spinors in the {\bf 1} of $SO(2)$, with $\gamma_3\eta_1=\eta_1$ and $\gamma_3\eta_2=\mp\eta_2$ in IIA/IIB.

Plugging the above expressions for $\chi_{1,2}$ into eq.~(\ref{calform}, \ref{calfin}), taking eqs.~(\ref{bilins}, \ref{bilinstr}, \ref{bilinscomp}) into account,  we 
find that the only non-vanishing calibration forms occur at codimensions zero (spacetime-filling), two, three, four, six and seven with respect to the external spacetime. The explicit expressions read:
\eq{\spl{\label{cald8}
\omega^{(0)}&= e^{8A-\Phi}\mathrm{Re}\Psi_1\\
\omega^{(2)}&= e^{6A-\Phi}\mathrm{Im}\Psi_1\\
\omega^{(3)}&= e^{5A-\Phi}\mathrm{Im}(e^{i\varphi}\Psi_2)\\
\omega^{(4)}&= e^{4A-\Phi}\mathrm{Re}\Psi_1\\
\omega^{(6)}&= e^{2A-\Phi}\mathrm{Im}\Psi_1\\
\omega^{(7)}&= e^{A-\Phi}\mathrm{Im}(e^{i\xi}\Psi_2)
~,}}
where we have taken 
eq.~(\ref{purespinors}) into account and we have set:  $(\widetilde{\theta}\theta)=e^{i\xi}$.  Moreover, the phase $\varphi$ on the right hand side of the third line comes from the normalization:
\eq{\label{phased8}
\frac{1}{4!}(\widetilde{\theta}\sigma_{i_1\dots i_4}\theta)~\!\mathrm{d}x^{i_1}\wedge \dots\wedge \mathrm{d}x^{i_4}=e^{i\varphi}\mathrm{vol}_{\mathrm{sp}}^{(4)}
~,}
where as in eq.~(\ref{calfin}) $\mathrm{vol}_{\mathrm{sp}}^{(q-1)}$ is  the 
unwarped volume density along the $(q-1)$-dimensional external space that the brane fills.

Taking into account the fact that the generalized calibrations (\ref{cald8}) obey the differential equation (\ref{caldiff}), we obtain the following equations:
\eq{\spl{\label{susyd8}
\mathrm{d}_H\left( e^{8A-\Phi}\mathrm{Re}\Psi_1\right)&=F^{\mathrm{el}}\\
\mathrm{d}_H\left( e^{6A-\Phi}\mathrm{Im}\Psi_1\right)&=0\\
\mathrm{d}_H\left( e^{4A-\Phi}\mathrm{Re}\Psi_1\right)&=0\\
\mathrm{d}_H\left(e^{2A-\Phi}\mathrm{Im}\Psi_1\right)&=0\\
\mathrm{d}_H\left(e^{5A-\Phi}\Psi_2 \right)&=0\\
\mathrm{d}_H\left(e^{A-\Phi}\Psi_2 \right)&=0
~.}}
Note that the last two equations above are precisely the GCY condition for the pure spinors $e^{5A-\Phi}\Psi_2$, $e^{A-\Phi}\Psi_2$, respectively; they are obtained by imposing $\mathrm{d}_H\omega^{(3,7)}=0$ for all $\varphi, \xi$; with $\omega^{(3,7)}$ given in eq.~(\ref{cald8}).

Based on the results for $d=4,6$, we conjecture that the content of eqs.~(\ref{susyd8}) should be precisely equivalent to the supersymmetry equations for the background. Assuming the correspondence to be true, the results of this section are summarized in the $d=8$ column of table \ref{table} and eq.~(1.1) of the introduction.

\subsection{d=2}\label{d2}

Let us consider supersymmetric backgrounds of the form  $\mathbb{R}^{1,1}\times\cm_8$ preserving one complex supercharge, corresponding 
to $\mathcal N=(2,0)$ supersymmetry in two dimensions.

Under $SO(9)\rightarrow SO(8)$ the nine-dimensional gamma matrices and  charge conjugation matrix decompose as:
\eq{
\label{epsdecompd2}
\qquad \Gamma_{{m}}=
\gamma_m  \, ,\qquad\Gamma_{{9}} =  {\gamma}_{{9}} \, ,  \qquad C_{9} =  C_8~,
}
where $\left\{\gamma_m, ~m=1,\dots,8\right\}$ are eight-dimensional gamma matrices, and $\gamma_9$ is the eight-dimensional chirality matrix. 
It can then  be seen that the explicit form of the Killing spinor decomposition eq.~(\ref{spindecompc}) reads:
\eq{\label{zu}
\chi_i=\frac{1}{\sqrt{2}}\left(e^{\frac{i\varphi}{2}}\eta_i+e^{-\frac{i\varphi}{2}}\eta_i^c\right)
~,}
where $i=1,2$, $\varphi$ is a phase and $\eta_{1}$, $\eta_{2}$ are pure Weyl spinors\footnote{Note that, as already mentioned, not all Weyl spinors of $SO(8)$ are pure. If $\eta$ is a  pure Weyl spinor, the purity of $\eta$ can be seen to be equivalent to the condition:
\eq{\label{pure8}
\widetilde{\eta}\eta=0~.
}
This condition can only be satisfied (for non-vanishing $\eta$) if $\eta$ 
is complexified. Given a pair ${\eta}_R$, ${\eta}_I$ of orthogonal Majorana-Weyl spinors of $SO(8)$ of the same chirality:
\eq{
\widetilde{\eta}_R\eta_R=\widetilde{\eta}_I\eta_I=|a|^2~;
~~~~\widetilde{\eta}_R\eta_I=0
~,\nn}
one can construct the complexified pure spinor $\eta$ through:
\eq{
\eta:=\frac{1}{\sqrt{2}}(\eta_R+i\eta_I)
~.\nn}
These are precisely the conditions imposed on the internal part of the Killing spinor in $\mathcal{N}=2$ M-theory compactifications on eight-manifolds of the type considered in \cite{beck}.
} in the {\bf 8} of $SO(8)$, with $\gamma_9\eta_1=\eta_1$ and $\gamma_9\eta_2=\mp\eta_2$ in IIA/IIB.

Plugging the above expressions for $\chi_{1,2}$ into eq.~(\ref{calform}, \ref{calfin}), taking eqs.~(\ref{bilins}, \ref{bilinstr}, \ref{bilinscomp}) into account,  we 
find that the only non-vanishing calibration form occurs at codimensions zero (spacetime-filling). The explicit expression reads:
\eq{\spl{\label{cald2}
\omega^{(0)}&= e^{2A-\Phi}\mathrm{Re}\left(e^{i\varphi}\Psi_2\right)+e^{2A-\Phi}\mathrm{Re}\Psi_1
~,}}
where we have taken 
eq.~(\ref{purespinors}) into account. The phase $\varphi$ on the right hand side is the same as in eq.~(\ref{zu}).

Taking into account the fact that the generalized calibrations (\ref{cald2}) obey the differential equation (\ref{caldiff}), we obtain the following equations:
\eq{\spl{\label{susyd2}
\mathrm{d}_H\left( e^{2A-\Phi}\mathrm{Re}\Psi_1\right)&=F^{\mathrm{el}}\\
\mathrm{d}_H\left(e^{2A-\Phi}\Psi_2 \right)&=0
~.}}
The equations above are obtained by imposing $\mathrm{d}_H\omega^{(0)}=F^{\mathrm{el}}$ for all $\varphi$, with $\omega^{(0)}$ given in eq.~(\ref{cald2}). 
The second of the two equations is precisely the GCY condition for the pure spinor $e^{2A-\Phi}\Psi_2$.
\newpage

Based on the results for $d=4,6$, we conjecture that the content of eqs.~(\ref{susyd2}) should be precisely equivalent to the supersymmetry equations for the background. Assuming the correspondence to be true, the results of this section are summarized in the $d=2$ column of table \ref{table} and eq.~(1.1) of the introduction.

\section{Conclusions}\label{sec:conclusions}

We considered type II backgrounds of the form $\mathbb{R}^{1,d-1}\times\cm_{10-d}$ for even $d$, preserving 
$2^{d/2-1}$ complex supercharges -- as many as the components of 
a complexified Weyl spinor of $SO(1,d-1)$. 
For $d=6$ we proved that there is a one-to-one correspondence between 
background supersymmetry equations in pure-spinor form and D-brane 
generalized calibrations -- a fact which was already known in the $d=4$ case. 
We conjectured that this one-to-one correspondence should hold for general $d$, 
and used this to `predict' the background supersymmetry equations 
in pure-spinor form for the $d=2,8$ cases. It would be nice to verify our conjecture for $d=2,8$ by either a brute-force  computation, as we have done here in the $d=6$ case, or by a counting argument, as in \cite{scan} for the $d=4$ case. 

We expect our results for the background supersymmetry equations in pure-spinor form to be useful in finding novel flux vacua. Of course in this case one would also have to solve the Bianchi identities in addition to the supersymmetry equations \cite{lt, gaunteqs, kt}. The study of generalized calibrations has shed light to the construction of effective actions, and has recently suggested a way to break supersymmetry in a controlled way \cite{march,held}. 
It would be interesting to pursue this connection further. It would also be interesting to repeat our analysis for backgrounds of the form AdS$_d\times\cm_{10-d}$, along the lines of \cite{koermart}.

\section*{Acknowledgement}

We would like to thank Diederik Roest for useful discussions.

\appendix

\section{Spinors and gamma matrices in Euclidean spaces}\label{sec:spinors}

In this section we list some useful relations and 
explain in more detail our spinor conventions for general even-dimensional Euclidean spaces of dimension $2k$. 

The charge conjugation matrix obeys:
\eq{\label{c}
C^{\mathrm{Tr}}=(-)^{\frac{1}{2}k(k+1)}C~;~~C^*=(-)^{\frac{1}{2}k(k+1)}C^{-1}~;~~
\gamma_m^{\mathrm{Tr}}=(-)^k C^{-1}\gamma_m C
~.
}
The complex conjugate $\eta^c$ of a spinor $\eta$ is given by:
\eq{
\eta^c:=C\eta^*
~,}
form which it follows that:
\eq{
(\eta^c)^c=(-)^{\frac12 k(k+1)}\eta
~.}
The chirality matrix $\gamma_{2k+1}$ is defined by:
\eq{\label{chirdef}
\gamma_{2k+1}:=i^k\gamma_1\dots \gamma_{2k}
~,}
and obeys
\eq{\label{chirob}
\gamma^{\mathrm{Tr}}_{2k+1}
=(-)^k C^{-1}\gamma_{2k+1}C
~,}
as follows from eqs.~(\ref{chirdef}, \ref{c}). The chirality projector: 
\eq{
P_{\pm}:=\frac12(1\pm\gamma_{2k+1})
~,} 
projects a Dirac spinor $\chi$ onto the chiral, antichiral Weyl parts $\chi_\pm$:
\eq{
\chi_\pm=P_\pm\chi~.
}
Taking eq.~(\ref{chirob}) into account we obtain:
\eq{\label{ptr}
C^{-1}P_\pm=
\left\{
\begin{array}{ll}
P_\pm^{\mathrm{Tr}}C^{-1}~,& ~~~k=\mathrm{even}\\
P_\mp^{\mathrm{Tr}}C^{-1}~,& ~~~k=\mathrm{odd}\\
\end{array}
\right.
~.}
Covariantly-transforming spinor bilinears must be of the form $(\widetilde{\psi}\gamma_{m_1\dots m_p}\chi)$, 
where in any dimension we define:
\eq{
\widetilde{\psi}:=\psi^{\mathrm{Tr}}C^{-1}
~.}
Using eq.~(\ref{ptr}) we find:
\eq{\spl{\label{bilins}
(\widetilde{\psi}_\pm\gamma_{m_1\dots m_{2l}}\chi_\mp)&=0=(\widetilde{\psi}_\pm\gamma_{m_1\dots m_{2l+1}}\chi_\pm)~,~~k=\mathrm{even}\\
(\widetilde{\psi}_\pm\gamma_{m_1\dots m_{2l}}\chi_\pm)&=0=(\widetilde{\psi}_\pm\gamma_{m_1\dots m_{2l+1}}\chi_\mp)~,~~k=\mathrm{odd}
~.}}
Moreover:
\eq{\label{bilinstr}
(\widetilde{\psi}\gamma_{m_1\dots m_p}\chi)=
(-)^{kp+\frac12 k(k+1)}(\widetilde{\chi}\gamma_{m_p\dots m_1}\psi)
=
(-)^{\frac12 (k-p)(k-p+1)}(\widetilde{\chi}\gamma_{m_1\dots m_p}\psi)
~.}
The identity
\eq{
\gamma_{m_1\dots m_p}^*=(-)^{kp}C^{-1}\gamma_{m_1\dots m_p}C
~,}
can be used to show the following relations:
\eq{\spl{\label{bilinscomp}
(\widetilde{\psi}\gamma_{m_1\dots m_p}\chi)^*&
=(-)^{kp}(\widetilde{\psi^c}\gamma_{m_1\dots m_p}\chi^c)\\
(\widetilde{\psi}\gamma_{m_1\dots m_p}\chi^c)^*&
=(-)^{kp+\frac12 k(k+1)}(\widetilde{\psi^c}\gamma_{m_1\dots m_p}\chi)
~.}}

\section{Explicit solution of the supersymmetry equations in $d=6$}\label{sec:explicit}

In this section we give the details of the derivation of the explicit solution of the Killing spinor equations for type II $\mathbb{R}^{1,5}\times\cm_4$ 
flux backgrounds with minimal supersymmetry in six dimensions. The IIA, IIB cases are treated separately in sections \ref{iia}, \ref{iib} below. The requirement that the background admits a pair of globally-defined, nowhere-vanishing pure spinors, leads to a different topological condition in each case: on the type IIA side it implies the trivialization of the structure group of $T\cm_4$, while on the IIB side it implies the reduction of the structure group of $T\cm_4$ to $SU(2)$; see \cite{trendl,triendlthesis} for a recent discussion. 

The explicit solution of the Killing spinor equations ({\it i.e.} the supersymmetry conditions) given below can be seen to be identical to the solution of the set of pure-spinor equations (\ref{susyd6}) of section \ref{d6} -- which are the differential conditions obeyed by static, magnetic D-brane calibrations. Thus we provide here for the $d=6$ case a proof of the one-to-one correspondence between background supersymmetry pure-spinor equations and D-brane calibrations.

Our starting point is the ten-dimensional supersymmetry equations:
\begin{align}\label{killingspinoreqs}\begin{split}
0&=\left(\slashed{\partial}\Phi+\frac{1}{2}\slashed{H}\right)\epsilon_1
+\left(\frac{1}{16}e^\Phi\Gamma^M\slashed{F}\Gamma_M\Gamma_{11}\right)\epsilon_2\\
0&=\left(\slashed{\partial}\Phi-\frac{1}{2}\slashed{H}\right)\epsilon_1
-\left(\frac{1}{16}e^\Phi\Gamma^M\sigma(\slashed{F})\Gamma_M\Gamma_{11}\right)\epsilon_1\\
0&=\left(\nabla_M+\frac{1}{4}\slashed{H}_M\right)\epsilon_1
+\left(\frac{1}{16}e^\Phi\slashed{F}\Gamma_M\Gamma_{11}\right)\epsilon_2\\
0&=\left(\nabla_M-\frac{1}{4}\slashed{H}_M\right)\epsilon_2-\left(\frac{1}{16}e^\Phi\sigma(\slashed{F})\Gamma_M\Gamma_{11}\right)\epsilon_1
~,
\end{split}\end{align}
where the ten-dimensional Majorana-Weyl Killing spinors are decomposed 
as in (\ref{spindecompa}). The ten-dimensional gamma matrices $\Gamma_M$ are decomposed as follows:
\eq{
\label{gammadecompapp}
\qquad \Gamma_{{\mu}}=
\hat{\gamma}_{\mu} \otimes \bbone  \, ,\qquad\Gamma_{{m+5}} = \hat{\gamma}_7 \otimes {\gamma}_{{m}} ~,
}
where $\left\{\hat{\gamma}_{\mu}, ~\mu=0,\dots,5\right\}$, $\left\{\gamma_m, ~m=1,\dots,4\right\}$ are six-, four-dimensional gamma matrices, respectively, and $\hat{\gamma}_7$ is the six-dimensional chirality matrix. 
In the following subsections we will consider the IIA, IIB cases separately.

\subsection{IIA}\label{iia}

We may parameterize the internal nowhere-vanishing, globally-defined Weyl spinors $\eta_{1,2}$ in the Killing-spinor ansatz in (\ref{spindecompa}) 
as follows:
\eq{\label{etachidef}
\eta_1=a~\!\eta~,~~~ \eta_2=b~\! \chi
~,
}
where $\eta$, $\chi$ are unimodular Weyl spinors of 
opposite chirality and $|a|^2=|b|^2$. Moreover we can choose without loss of generality the phases of  $\eta$, $\chi$ so that $a=b\in\mathbb{R}$.

The pair of nowhere-vanishing, globally-defined Weyl spinors  $\eta$, $\chi$ trivializes the tangent bundle of $\cm_4$, so that the structure group reduces to $\bbone$. This can also be seen by constructing a pair of complex vectors:
\begin{equation}\label{uvdef}
u^m=\widetilde{\eta}\gamma^m\chi~;~~~
v^m=\widetilde{\eta}\gamma^m\chi^c
~.
\end{equation}
As can be proven by Fierzing, the four real globally-defined vectors $\mathrm{Re}u$, $\mathrm{Im}u$, $\mathrm{Re}v$, $\mathrm{Im}v$ are unimodular and mutually orthogonal; hence they provide an explicit trivialization of 
the tangent bundle $T\cm_4$. 

Let us also mention that in deriving the general solution to the Killing spinor equations, it will be  useful to take the following  relations into account:
\begin{equation}\label{useful}\begin{split}
\gamma_m\eta&=v_m\chi-u_m\chi^c\\
\gamma_m\eta^c&=v_{m}^*\chi^c+u_{m}^*\chi\\
\gamma_m\chi&=v_{m}^*\eta+u_m\eta^c\\
\gamma_m\chi^c&=v_m\eta^c-u_{m}^*\eta
~,
\end{split}\end{equation}
which can be shown by Fierzing. 

We now proceed by decomposing all forms on the basis of $u$, $v$ -- which can also be thought of as one-forms given the existence of a metric on $\cm_4$; in the following we will use the same notation for  both the vectors and the one forms.

The most general decomposition of the various components of the (magnetic) RR flux $F$, {\it cf.} eq.~(\ref{fluxan}), reads as follows:
\eq{F=F_0+F_2+F_4~,
}
with:
\begin{equation}\begin{split}\label{fluxiia}
e^\Phi F_0&=f^{(0)}\\
e^\Phi F_2&=\frac{1}{2}(\i f^{(2)}_1 u\wedge u^*+\i f^{(2)}_2 v\wedge v^*+f^{(2)}_3 u\wedge v+f^{(2)*}_3 u^*\wedge v^*+f^{(2)}_4 u\wedge v^*+f^{(2)*}_4 u^*\wedge v)\\
e^\Phi F_4&=\frac{1}{4}f^{(4)} u\wedge v\wedge u^*\wedge v^*
~,
\end{split}\end{equation}
where $f^{(0)}$, $f^{(2)}_{1,2}$, $f^{(4)}$ are real scalars, and  
$f^{(2)}_{3,4}$ are complex scalars. Similarly, we decompose the NSNS three-form  as follows:
\begin{equation}{H}=\star_4\left(h_1 u+h_2 v+\mathrm{c.c.}\right)
~,
\end{equation}
where $h_{1,2}$ are complex scalars.

We also need the decompositions of the derivatives of the real scalars $\Phi$, $A$, $a$:
\begin{equation}\begin{split}\label{derivsiia}
\partial_m\Phi&=\tfrac{1}{2}\left(u^*_m\varphi_u+v^*_m\varphi_v+\mathrm{c.c.}\right)\\
\partial_mA&=\tfrac{1}{2}\left(u^*_mA_u+v^*_mA_v+\mathrm{c.c.}\right)\\
\partial_m a&=\tfrac{1}{2}\left(u^*_m(\partial a)_u+v^*_m(\partial a)_v+\mathrm{c.c.}\right)~,
\end{split}\end{equation}
where $\varphi_u$, $\varphi_v$, $A_u$, $A_v$, $(\partial a)_u$, $(\partial a)_v$, are complex scalars.

The torsion classes of the (trivial) structure of $T\cm_4$ parameterize the failure of $\eta$, $\chi$ to be covariantly constant. Explicitly, we define 
the torsion classes $\mathcal{W}_m^{(i)}$, $i=1,\dots 4$, via:
\begin{equation}\begin{split}\label{torsioniia}
\nabla_m\eta&=\mathcal{W}_m^{(1)}\eta+\mathcal{W}_m^{(2)}\eta^c\\
\nabla_m\chi&=\mathcal{W}_m^{(3)}\chi+\mathcal{W}_m^{(4)}\chi^c
~,
\end{split}\end{equation}
where $\mathcal{W}^{(2,4)}$ are complex one-forms, and $\mathcal{W}^{(1,3)}$ are imaginary one-forms; the latter property  follows from the definition (\ref{torsioniia}) upon taking the unimodularity of $\eta$, $\chi$ into account. Explicitly, for $i=1,\dots,4$ we decompose:
\begin{equation}\begin{split}
\mathcal{W}^{(i)}=\tfrac{1}{2}(u^*\mathcal{W}^{(i)}_u
+v^*\mathcal{W}^{(i)}_v+u\mathcal{W}^{(i)}_{u^*}+v\mathcal{W}^{(i)}_{v^*})
~,
\end{split}\end{equation}
where $\mathcal{W}^{(i)}_u$ $\mathcal{W}^{(i)}_v$, $\mathcal{W}^{(i)}_{u^*}$,  $\mathcal{W}^{(i)}_{v^*}$ are complex scalars. Moreover, the fact that $\mathcal{W}^{(1,3)}$ are imaginary implies:
\eq{
\mathcal{W}^{(i)}_{u^*}=-\mathcal{W}^{(i)}_u~;~~~
\mathcal{W}^{(i)}_{v^*}=-\mathcal{W}^{(i)}_v
~,}
for $i=1,3$. Let us also note that alternatively the torsion classes can 
be defined in terms of the exterior derivatives of $u$, $v$. Indeed, from
 eq.~(\ref{torsioniia}) we have, upon taking definition (\ref{uvdef}) into account:
\begin{equation}\begin{split}\label{torsioniiaalt}
\d u&=(\mathcal{W}^{(1)}+\mathcal{W}^{(3)})\wedge u+\mathcal{W}^{(4)}\wedge v-\mathcal{W}^{(2)}\wedge v^*\\
\d v&=(\mathcal{W}^{(1)}-\mathcal{W}^{(3)})\wedge v-\mathcal{W}^{(4)*}\wedge u+\mathcal{W}^{(2)}\wedge u^*~.
\end{split}\end{equation}

We are now ready to give the general solution to the background supersymmetry equations, by plugging the above expansions into the Killing spinor equations (\ref{killingspinoreqs}), taking eq.~(\ref{useful}) into account. The solution is parameterized in terms of  eight real unconstrained scalar degrees of freedom which we may take to be $f^{(0)}$, $f^{(2)}_1$, $f^{(2)}_4$, $\varphi_u$, $\varphi_v$ (recall that the last two complex scalars parameterize $\partial_m\Phi$). Explicitly, the torsion classes are given by:
\begin{equation}\begin{split}
W^{(1)}_u&=\frac{1}{2}\varphi_u-f^{(2)*}_4\\
W^{(1)}_v&=\frac{1}{2}\varphi_v-\frac{\i}{4}f^{(2)}_1-\frac{1}{2}f^{(0)}\\[0.5cm]
W^{(2)}_u&=0\\
W^{(2)}_v&=-\frac{1}{2}f^{(2)*}_4\\
W^{(2)}_{u^*}&=\varphi_v-\i f^{(2)}_1-f^{(0)}\\
W^{(2)}_{v^*}&=-\varphi_u+\frac{3}{2}f^{(2)*}_4\\[0.5cm]
W^{(3)}_u&=\frac{1}{2}\varphi_u-f^{(2)*}_4\\
W^{(3)}_v&=-\frac{1}{2}\varphi_v+\frac{\i}{4}f^{(2)}_1+\frac{1}{2}f^{(0)}\\[0.5cm]
W^{(4)}_u&=0\\
W^{(4)}_v&=\varphi_u-\frac{3}{2}f^{(2)*}_4\\
W^{(4)}_{u^*}&=-\varphi_v^*-\i f^{(2)}_1+f^{(0)}\\
W^{(4)}_{v^*}&=\frac{1}{2}f^{(2)*}_4~.
\end{split}\end{equation}
The NSNS and RR fluxes are given by:
\begin{equation}\begin{split}
h_1&=-\varphi_u^*+\frac{3}{2}f^{(2)}_4\\
h_2&=-\varphi_v^*-\frac{3\i}{4}f^{(2)}_1+\frac{5}{4}f^{(0)}\\[0.5cm]
f^{(2)}_2&=0\\
f^{(2)}_3&=-f^{(2)}_4\\[0.5cm]
f^{(4)}&=0
~.
\end{split}\end{equation}
Finally, the derivatives of $a$, $A$, {\it cf.} eq.~(\ref{derivsiia}), read:
\begin{equation}\begin{split}
(\partial a)_u&=\frac{a}{4}f^{(2)*}_4\\
(\partial a)_v&=\frac{\i a}{8}f^{(2)}_1+\frac{a}{8}f^{(0)}
\end{split}\end{equation}
and
\begin{equation}\begin{split}
A_u&=\frac{1}{2}f^{(2)*}_4\\
A_v&=\frac{\i}{4}f^{(2)}_1+\frac{1}{4}f^{(0)}~.
\end{split}\end{equation}
We therefore see explicitly that $|a|^2\propto e^A$, as already mentioned above eq.~(\ref{ea}).

To make contact with the supersymmetry equations in pure-spinor form (\ref{susyd6}), we note that the definition (\ref{purespinors}) implies:
\begin{equation}\begin{split}\label{purespinsiia}
\Psi_1&=v-\frac{1}{2}u\wedge v\wedge u^*\\
\Psi_2&=u+\frac{1}{2}u\wedge v\wedge v^*
~,
\end{split}\end{equation}
where we have taken eqs.~(\ref{etachidef}), (\ref{uvdef}), (\ref{useful}) into account. It is then straightforward to show that the solution of the Killing spinor equations given above is identical to the solution one obtains by substituting (\ref{purespinsiia}) into eqs.~(\ref{susyd6}), taking (\ref{fluxiia}) - (\ref{derivsiia}), (\ref{torsioniiaalt}) into account.

\subsection{IIB}\label{iib}


We parameterize the internal nowhere-vanishing, globally-defined Weyl spinors $\eta_{1,2}$ in the Killing-spinor ansatz in (\ref{spindecompa}) 
as follows:
\eq{\label{etachidefiib}
\eta_1=a~\!\eta~,~~~ \eta_2=b~\! \eta +c~\! \eta^c
~,
}
where $\eta$ is a unimodular Weyl spinor of 
positive chirality. Without loss of generality 
we may choose the phase of $\eta$  so that $a\in\mathbb{R}$; the 
scalars $b$, $c$ are in general complex.

The nowhere-vanishing, globally-defined Weyl spinor  $\eta$ reduces the structure group of the tangent bundle of $\cm_4$ to $SU(2)$. This can also be seen by constructing a real two-form $j$ and a complex two-form $\omega$ on $\cm_4$ as spinor bilinears:
\begin{equation}\begin{aligned} \label{jomegadef}
j_{mn}&=\i\tilde{\eta}\gamma_{mn}\eta^c&\qquad\omega_{mn}&=-\i\tilde{\eta}\gamma_{mn}\eta
~.
\end{aligned}\end{equation}
The pair $(j,\omega)$ defined above, can be seen by Fierzing to obey the 
definition of an $SU(2)$ structure:
\eq{
j\wedge\omega=0~;~~~j\wedge j=\frac12 \omega\wedge\omega^*\neq0
~.}

On $\cm_4$ there is an almost complex structure, which can be given explicitly in terms of the projectors:
\eq{
\left(\Pi^\pm\right)_m{}^n:=\frac12\left( \delta_m{}^n\mp i j_m{}^n\right)
~.}
A one-form $V$ can thus be decomposed into (1,0) and  (0,1) parts $V^+$, $V^-$ with respect to the almost complex structure via: $V_m^\pm:=\left(\Pi^\pm\right)_m{}^n V^{\ }_n$. 
We will also make use of the following definitions:
\begin{equation}\begin{split}
\widetilde{V}_m^-&=\tfrac{\i}{2}\omega_{mn}^*V^{n+}\\
\widetilde{V}_m^+&=-\tfrac{\i}{2}\omega_{mn}V^{n-}
~,
\end{split}\end{equation}
for any real vector $V_m$. 
Let us also mention that in deriving the general solution to the Killing spinor equations, it will be useful to take the following  relations into account:
\eq{\spl{\label{usefuliib}
\gamma_{mn}\eta&=ij_{mn}\eta+i\omega_{mn}\eta^c\\
\gamma_{mn}\eta^c&=-ij_{mn}\eta^c+i\omega_{mn}^*\eta
~,}}
which can be shown by Fierzing; see \cite{gauntlett, blt, trendl, triendlthesis} 
for a more detailed discussion of $SU(2)$ structures.

The torsion classes of the $SU(2)$ structure of $T\cm_4$ parameterize the failure of $\eta$ to be covariantly constant. Explicitly, we define 
the torsion classes $\mathcal{W}_m^{(i)}$, $i=1,2$, via:
\begin{equation}\begin{split}\label{torsioniib}
\nabla_m\eta&=\mathcal{W}_m^{(1)}\eta+\mathcal{W}_m^{(2)}\eta^c
~,
\end{split}\end{equation}
where as in the IIA case, $\mathcal{W}^{(2)}$ is a complex one-form, and $\mathcal{W}^{(1)}$ is an imaginary one-form. Alternatively the torsion classes can be defined in terms of the exterior derivatives of $j$, $\omega$. 
Indeed, from eq.~(\ref{torsioniib}) we have, upon taking definition (\ref{jomegadef}) into account:
\eq{\spl{\label{torsioniibalt}
\d j &= \mathcal{W}_2^*\wedge\omega+\mathcal{W}_2\wedge\omega^* \\
\d \omega &=2\mathcal{W}_1\wedge\omega-2\mathcal{W}_2\wedge j~.
}}

As already mentioned, the spinor $\eta$ further reduces the structure group of $T\cm_4$ from  $SO(4)\cong SU(2)\times SU(2)'$ (which is accomplished by the 
existence of a Riemannian metric on $\cm_4$) to $SU(2)$. The spinors $\eta$, $\eta^c$ are singlets under the first $SU(2)$ factor, whereas they transform as an $SU(2)'$ doublet under the second factor. Moreover there is an alternative $SU(2)'$-covariant description of the $SU(2)$ structure on $T\cm_4$ and its associated torsion classes, which can be seen as follows:\footnote{The following two equations were worked out together with Diederik Roest.} Let us define a  triplet of real two-forms $j_i$, and a triplet of 
real one-forms $\mathcal{W}_i$, $i=1,2,3$, via 
\eq{
(j_1,j_2,j_3):=(j,\mathrm{Re}\omega,
-\mathrm{Im}\omega)
~;~~~
(\mathcal{W}_1,\mathcal{W}_2,\mathcal{W}_3):=(\mathrm{Im}\mathcal{W}^{(1)},\mathrm{Im}\mathcal{W}^{(2)},
-\mathrm{Re}\mathcal{W}^{(2)})
~.}
It can be seen that the $j_i$'s transform as a triplet of $SU(2)'$, and moreover eqs.~(\ref{torsioniibalt}) can be cast in an $SU(2)'$-covariant form:
\eq{
\d j_m=2\varepsilon_{mnp}\mathcal{W}_n \wedge j_p
~.}
We may use this $SU(2)'$ gauge freedom to rotate the torsion classes in eq.~(\ref{torsioniibalt}) to a more standard form, as in \cite{gauntlett}.

We now proceed by giving the most general ansatz for  all forms. 
The (magnetic) RR flux $F$, {\it cf.} eq.~(\ref{fluxan}), can be expanded as:
\begin{equation}{F}=F_1+F_3~,\end{equation}
with:
\begin{equation}\begin{split}\label{fluxiib}
e^\Phi F_1&=f^{(1)}\\
e^\Phi F_3&=\star_4 f^{(3)}
~,
\end{split}\end{equation}
where $f^{(1)}$, $f^{(3)}$, are real one-forms on $\cm_4$. Similarly, we decompose the NSNS three-form  as follows:
\begin{equation}\label{hiib}
{H}=\star_4h
~,
\end{equation}
where $h$ is a real one-form.

We are now ready to give the solution to the background supersymmetry equations, by plugging the above expansions into the Killing spinor equations (\ref{killingspinoreqs}), taking eq.~(\ref{usefuliib}) into account.  The equality of the norms of $\eta_{1,2}$ imposes:
\eq{a^2=|b|^2+|c|^2
~,}
which we will assume to hold in the following; moreover, we will take $b\in\mathbb{R}$ for simplicity. 
Explicitly, the fluxes are given by:
\begin{equation}\begin{split}
f_m^{(1)}&=-\frac{4}{a}\left(c(\widetilde{\partial_mA})^-+c^*(\widetilde{\partial_mA})^+\right)\\
f_m^{(3)}&=\frac{4b}{a}(\partial_mA)\\[0.5cm]
h_m&=-\frac{4b}{a^2}\left(c(\widetilde{\partial_mA})^-+c^*(\widetilde{\partial_mA})^+\right)
~.
\end{split}\end{equation}
The torsion classes read:
\begin{equation}\begin{split}
\mathcal{W}_m^{(1)+}&=\frac{2b^2-a^2}{2a^2}(\partial_mA)^+\\
\mathcal{W}_m^{(1)-}&=-\frac{2b^2-a^2}{2a^2}(\partial_mA)^-\\[0.5cm]
\mathcal{W}_m^{(2)+}&=\frac{b}{a^2}\left(c(\partial_mA)^++b(\widetilde{\partial_mA})^+\right)\\
\mathcal{W}_m^{(2)-}&=\frac{c}{a^2}\left(c(\widetilde{\partial_mA})^-+b(\partial_mA)^-\right)~.
\end{split}\end{equation}
Moreover, we have:
\begin{equation}\begin{split}
(\partial_ma)&=\frac{a}{2}(\partial_mA)\\
(\partial_mb)&=\frac{1}{2}b\frac{5a^2-4b^2}{a^2}(\partial_mA)\\
(\partial_mc)&=c\frac{a^2-4b^2}{2a^2}(\partial_mA)
\end{split}\end{equation}
and
\begin{equation}\begin{split}
(\partial_m\Phi)&=\frac{2(a^2+|c|^2)}{a^2}(\partial_mA)^-~.
\end{split}\end{equation} 

To make contact with the supersymmetry equations in pure-spinor form (\ref{susyd6}), we note that the definition (\ref{purespinors}) implies:
\begin{equation}\begin{split}\label{purespinsiib}
\Psi_1&=\frac{1}{a}\left(b-b\text{vol}_4-\i(bj+c^*\omega)\right)\\
\Psi_2&=\frac{1}{a}\left(c-c\text{vol}_4+\i(b\omega-cj)\right)
~,
\end{split}\end{equation}
where $\text{vol}_4$ is the volume form of $\cm_4$, and we have taken eqs.~(\ref{etachidefiib}), (\ref{jomegadef}), (\ref{usefuliib}) into account. It is then straightforward to show that the solution of the Killing spinor equations given above also solves eqs.~(\ref{susyd6}), upon  taking (\ref{purespinsiib}), (\ref{torsioniibalt}), (\ref{fluxiib}), (\ref{hiib}) into account.

\section{Explicit examples}\label{sec:warpedk3}

As an illustration of the pure-spinor formalism in the $d=6$ case, we will now construct a type IIB warped K3 solution with spacetime-filling D5 branes  localized on the K3. We also construct a IIA warped $S^1\times T^3$ solution with spacetime-filling D6 branes localized on the $T^3$ and wrapping the $S^1$.
The two solutions are related by T-duality, in the case where on the IIB side the K3 is replaced by a $T^4$.

\subsection*{The IIB solution}

The ten-dimensional metric is of the form:
\eq{
\d s^2=e^{2A}\d s^2(\mathbb{R}^{1,5})+e^{-2A}\d s^2(\mathrm{K3})
~.}
Correspondingly, the $SU(2)$ structure $(j,\omega)$ obeys:
\eq{\d (e^{2A}j)=\d (e^{2A}\omega)=0
~.}
Taking the equations above into account, it can be seen that 
the pure spinors of eq.~(\ref{purespinsiib}) solve the supersymmetry 
equations (\ref{susyd6}), provided the remaining fields are given by:
\begin{equation}\begin{split}
F_1&=0\\
F_3&=4e^{-2A}\star_4\d A\\
H&=0\\
\Phi&=2A~,
\end{split}\end{equation}
and we also set: $a=b$, $c=0$. 
In order to have a solution to the full set of equations of motion, it suffices to  impose in addition the Bianchi equations for all fields \cite{lt,gaunteqs,kt}. It is not difficult to see that this leads to one additional equation:
\eq{
\nabla^2_{\mathrm{K3}}e^{-4A}=0
~,}
{\it i.e.} $e^{-4A}$ is harmonic with respect to the K3 metric.\footnote{
A harmonic function on a smooth Riemannian manifold without boundary is constant; in our case this would lead to a constant warp factor, and all flux would vanish.} The solution also admits spacetime-filling D5 branes localized on K3 (as can be seen from the form of the RR three-form flux in the solution above), which can be introduced by replacing the right-hand side above with a delta function on K3. 

Upon replacing the K3 by a $T^4$,  the solution coincides with the one obtained using the `harmonic superposition rules' for a stack of D5 branes in flat space (see \cite{youm} for a review). Moreover, one can `smear' the warp factor $A$ along one direction of the torus ({\it i.e.} assume that $A$ is independent of the corresponding coordinate) and T-dualize to IIA along the smeared direction. The T-dual is a warped $S^1\times T^3$ solution with spacetime-filling D6 branes localized on the $T^3$ and wrapping the $S^1$.

\subsection*{The IIA solution}

We would now like to describe the T-dual warped $S^1\times T^3$ solution with spacetime-filling D6 branes, mentioned in the previous subsection,  in the language of pure spinors.

The ten-dimensional metric is of the form:
\eq{
ds^2=e^{2A}\left(\d s^2(\mathbb{R}^{1,5})+\d\lambda^2\right)+e^{-2A}\d s^2(T^3)
~,}
where the coordinate $\lambda$ parameterizes an $S^1$. 
Correspondingly, the complex one-forms $u$, $v$ may be chosen as follows:
\eq{
u=e^{-A}\left(\d y_1^2+\i \d y_2^2\right)~;~~~
v=e^{A}\d\lambda^2+\i e^{-A}\d y_3^2
~,}
where $y_1,y_2,y_3$ are coordinates of $T^3$ such that $\d{}s^2(T^3)=\d y_i\d y_i$. 
Taking the equations above into account, it can be seen that 
the pure spinors of eq.~(\ref{purespinsiia}) solve the supersymmetry 
equations (\ref{susyd6}), provided the remaining fields are given by:
\begin{equation}\begin{split}
F_0&=0\\
F_2&=-4e^{-2A}\star_4(\d A\wedge\d\lambda)\\
F_4&=0\\
H&=0\\
\Phi&=3A~.
\end{split}\end{equation}
In order to have a solution to the full set of equations of motion, it suffices to  impose in addition the Bianchi equations for all fields \cite{lt,gaunteqs,kt}. It is not difficult to see that this leads to one additional equation:
\eq{
\nabla^2_{\mathrm{T^3}}e^{-4A}=0
~,}
{\it i.e.} $e^{-4A}$ is harmonic with respect to the metric on $T^3$ ({\it cf.} the last footnote). The solution also admits spacetime-filling D6 branes localized on $T^3$ 
and wrapping the $S^1$ parameterized by $\lambda$ (as can be seen from the form of the RR two-form flux in the solution above), which can be introduced by replacing the right-hand side above with a delta function on $T^3$.

%
%

\bibliographystyle{JHEP}
\bibliography{literature}

\end{document}